\documentclass[11pt,article]{aastex}
\usepackage{emulateapj5}
\usepackage{graphicx}
\usepackage{grffile}
\usepackage{natbib}

\shorttitle{The First Planets}
\shortauthors{Johnson \& Li}

\begin{document}

\title{The First Planets: The Critical Metallicity for Planet Formation} 

\author{Jarrett L. Johnson\altaffilmark{1,2} and Hui Li\altaffilmark{1}}

\affil{$^{1}$Los Alamos National Laboratory, 
Los Alamos, NM  87545, USA  \\
Nuclear and Particle Physics, Astrophysics and Cosmology Group (T-2)}
\affil{$^{2}$Max-Planck-Institut f{\"u}r extraterrestrische Physik, 
Giessenbachstra\ss{}e, 85748 Garching, Germany \\
Theoretical Modeling of Cosmic Structures Group}

\email{jlj@lanl.gov}

\topmargin-0.2cm

\begin{abstract}
A rapidly growing body of observational results suggests that planet formation takes place preferentially at high metallicity. In the core accretion model of planet formation this is expected because heavy elements are needed to form the dust grains which settle into the midplane of the protoplanetary disk and coagulate to form the planetesimals from which planetary cores are assembled.  As well, there is observational evidence that the lifetimes of circumstellar disks are shorter at lower metallicities, likely due to greater susceptibility to photoevaporation.  Here we estimate the minimum metallicity for planet formation, by comparing the timescale for dust grain growth and settling to that for disk photoevaporation.  For a wide range of circumstellar disk models and dust grain properties, we find that the critical metallicity above which planets can form is a function of the distance $r$ at which the planet orbits its host star.  With the iron abundance relative to that of the Sun [Fe/H] as a proxy for the metallicity, we estimate a lower limit for the critical abundance for planet formation of [Fe/H]$_{\rm crit}$ $\simeq$ -1.5 + log($r$/1 AU), where an astronomical unit (AU) is the distance between the Earth and the Sun. This prediction is in agreement with the available observational data, and carries implications for the properties of the first planets and for the emergence of life in the early Universe.  In particular, it implies that the first Earth-like planets likely formed from circumstellar disks with metallicities $Z$ $\ga$ 0.1 Z$_{\odot}$.  If planets are found to orbit stars with metallicities below the critical metallicity, this may be a strong challenge to the core accretion model.  
\end{abstract}


\section{Introduction}
Following the formation of the first stars and galaxies, which transformed the Universe by bringing an end to the cosmic dark ages (e.g. Barkana \& Loeb 2001; Bromm \& Yoshida 2011), the
formation of the first planets marked another important transition in cosmic history.  Assembled from the heavy elements produced in the cores and supernovae of the first generations of 
stars, the first planets represent a milestone in the increasing complexity of the early Universe and set the stage for the emergence of the first life.    
When and where the first planets formed is therefore a question with bearing on topics ranging from early structure formation to the search for extraterrestrial intelligence.

A successful theory of planet formation should allow to make predictions of the properties of the earliest planets and their host stars which can be tested by observations of planetary systems
that may still be in the Galaxy today.  The popular core accretion model of planet formation (e.g. Pollack et al. 1996; Papaloizou \& Terquem 2006; Udry \& Santos 2007; Boley 2009; Janson et al. 2011), in particular, 
predicts that planets form first via the coagulation of dust grains 
into planetesimals, which are then assembled into planets with solid cores and in many cases gaseous atmospheres.  Dust being composed of elements heavier than the 
hydrogen, helium, and lithium forged in the Big Bang, this theory therefore demands that the first planets must have formed in protostellar disks already enriched by the metals ejected in the first supernovae.  
Indeed, it has long been expected that the first low-mass stars were also formed from such pre-enriched gas (e.g. Bromm \& Loeb 2003; Frebel et al. 2007; Clark et al. 2008), and perhaps from gas cooled by the dust synthesized in early supernovae (see e.g. Schneider et al. 2006; Caffau et al. 2011, 2012).  The minimum, or 'critical', metallicity for low-mass star formation is hotly debated, but is widely believed to be in the 
range 10$^{-6}$ - 10$^{-3.5}$ Z$_{\odot}$ (see e.g. Jappsen et al. 2009; Frebel 2010; Schneider et al. 2011).  The corresponding critical metallicity for planet formation is still an open question.

As expected from theoretical considerations (e.g. Ida \& Lin 2004; Kornet et al. 2005; Rice \& Armitage 2005; Johansen et al. 2009; Mordasini et al. 2012), observations of planetary systems 
(e.g. Santos et al. 2001; Fischer \& Valenti 2005; Maldonado et al. 2012) have established a correlation between the metallicity of a star and the likelihood that it hosts a planet (e.g. Papaloizou \& Terquem 2006; Williams \& Cieza 2011), or at least a relatively massive gas giant (e.g. Mayor et al. 2011).  In the core accretion model, there are at least two metallicity-dependent processes which must occur in order for a planet to be assembled within a dusty circumstellar disk (e.g. Weidenschilling 1980; Armitage 2010):  first, dust
grains must coagulate and settle into the midplane of the disk; second, dust grains and larger solid bodies in the disk midplane must grow via merging and accretion to form planetesimals and finally full-fledged 
planets.  Both the former (e.g. Kornet et al. 2005; Johansen et al. 2009)  and the latter (e.g. Ida \& Lin 2004; Rice \& Armitage 2005; Mordasini et al. 2012) of these processes have been shown to occur more readily in metal-rich disks than in metal-poor disks.
As planet formation must occur within the limited lifetime 
of protostellar disks (e.g. Yasui et al. 2009; Ercolano \& Clarke 2010), which of these two processes sets the minimum metallicity required for planet formation likely reduces to the question of which of them is the slower.  As noted by Armitage (2010), planetesimal
formation within the midplane is believed to be a relatively rapid process once dust settling has occurred; however, the process of dust settling can take much longer, especially in disk with low dust-to-gas ratios (i.e. 
at low metallicities).  This suggests that the main bottleneck to planet formation at low metallicities, and so to the formation of the first planets, is the slow process of dust settling in the disks surrounding 
metal-poor stars.  While previous studies accounting for the speed of dust coagulation have found this process to be critical for the formation of planets at metallicities $\ga$ 0.2 Z$_{\odot}$ (e.g. Kornet et al. 2005; Johansen et al. 2009), recent discoveries of planets orbiting stars of significantly lower metallicity ($\le$ 0.1 Z$_{\odot}$) by Niedzielski et al. (2009) and Setiawan et al. (2010) demonstrate that the critical 
metallicity for planet formation is well below what previous works have discussed (see also e.g. Gonzalez et al. 2001; Lineweaver 2001; Zinnecker 2004; Pinotti et al. 2005).

In the present work, we formulate an estimate of the critical metallicity for planetesimal formation in the early, dust coagulation phase of the core accretion model of planet formation.\footnote{This early phase in the core accretion scenario is sometimes referred to as the 'planetesimal hypothesis' (e.g. Chambers 2001).  Strictly speaking, in the present work we only model this early stage of planetesimal formation.}  In the next Section, we discuss the 
timescales of relevance to the problem, namely that for dust settling and that for the dispersal of the disk by photoevaporation.  In Section 3 we directly compare
these timescales to estimate the critical metallicity for planet formation as a function of the distance from the host star.  We then test this theoretical 
critical metallicity against the available observational data on planetary systems, including the Solar System, in Section 4.  We discuss the implications of our 
results for the first planets and life in Section 5, and for the core accretion model of planet formation in Section 6.  Finally, we close with our conclusions in Section 7.

\begin{figure*}[t]
  \centering
  \includegraphics[width=4.5in]{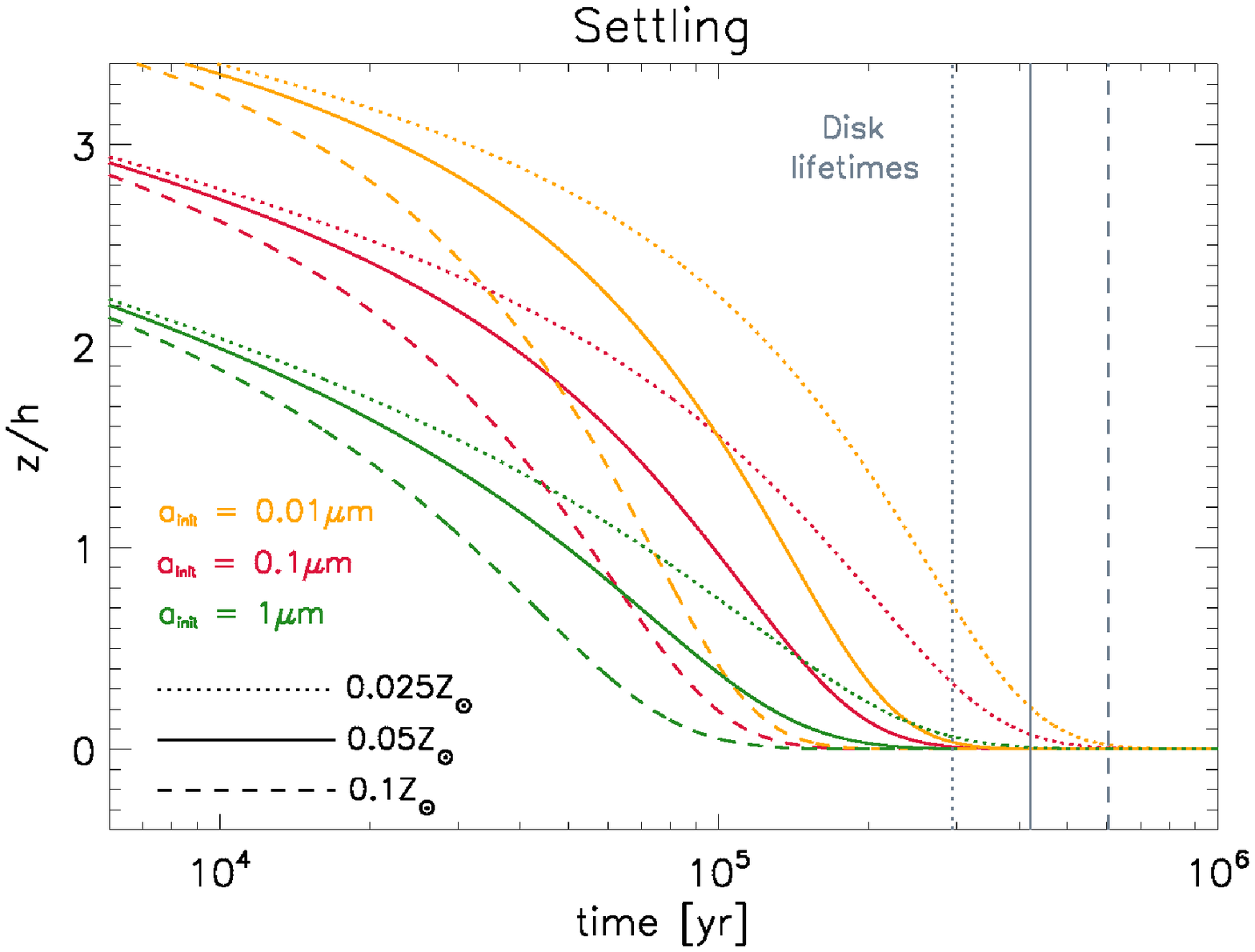}
  \caption
  {Dust grain settling timescales for initial grain sizes of 0.01 ({\it yellow}), 0.1 ({\it red}), and 1 $\mu$m ({\it green}), and for metallicities of 0.025 ({\it dotted lines}), 0.05 ({\it solid lines}), and 0.1 ({\it dashed lines}) Z$_{\odot}$.  Comparison of these timescales to the disk lifetimes at the same metallicities ({\it gray lines})
allows to estimate the minimum metallicity for which dust grains can settle into the equatorial region of the disk and there undergo runaway planetesimal formation before the disk is destroyed due to photoevaporation.  Dust grains can settle before disk photoevaporation at metallicities of $\ga$ 0.05 Z$_{\odot}$; however, even for the  largest initial sizes of $\sim$ 1 $\mu$m, they do not have time do so for a metallicity of 0.025 Z$_{\odot}$.  Therefore, for the disk surface density $\Sigma$ = 10$^3$ g cm$^{-2}$ and temperature $T$ = 200 K assumed here, the critical metallicity for planet formation is likely to lie within the range 0.025 - 0.05 Z$_{\odot}$, the exact value depending on the initial grain size distribution.}
\end{figure*}

\begin{figure*}[t]
  \centering
  \includegraphics[width=4.5in]{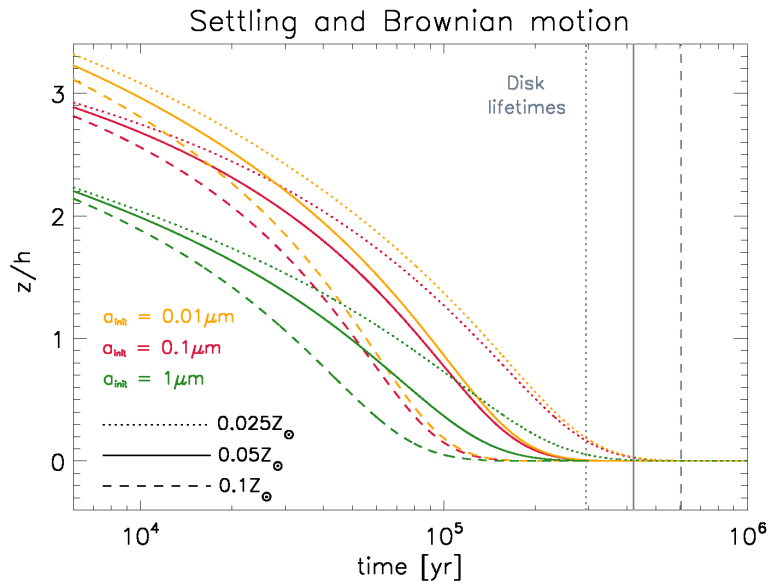}
  \caption
  {The same as Figure 1, but now for dust growth via both settling and Brownian motion.
As the timescale for growth via collisions due to Brownian motion is shorter for less massive grains (see equation 8), the smallest dust grains
experience growth and settling much faster in the calculation including Brownian motion than in the one neglecting it shown in Fig. 1.  For our calculation of the critical metallicity in Section 3, we include growth via both settling and Brownian motion.  }
\end{figure*}

\section{Time constraints on planet formation}
Here we describe two key timescales which set the critical metallicity
for planet formation.  The first is the time available for planet formation: the 
disk lifetime.  The second is the time required for dust grains to settle
into the midplane of the disk, where runaway planetesimal formation then takes place.

\subsection{Disk lifetime}
The two most widely discussed mechanisms of planetary disk dispersal are the formation of 
giant planets and photoevaporation by the host star.  One clue to the relative importance of each of these mechanisms comes from 
observations that have revealed low-metallicity disks to be more shortlived than those at higher metallicity (Yasui et al. 2009).
This is consistent with disk dispersal via photoevaporation being dominant, as disks at higher metallicity are more shielded
from the high energy radiation emitted from the central star (e.g. Ercolano \& Clarke 2010).  
Furthermore, the opposite trend could be expected for 
dispersal via giant planet formation, as giant planets tend to be more massive and to form more frequently in higher metallicity disks (e.g. Mordasini et al. 2012). 
For our estimate of the disk lifetime, we shall therefore assume that photoevaporation is the mechanism which dictates the lifetime of 
disks.  

Ercolano \& Clarke (2010) present calculations of the photoevaporation of planetary disks irradiated by X-ray emission from 
the central star.  These authors consider a wide range of disk metallicities, from 0.01 Z$_{\odot}$ to $\simeq$ 5 Z$_{\odot}$. 
They find higher X-ray photoevaporation rates at lower metallicity due to lower extinction within the disk, which allows gas 
at higher column densities to be photoheated and entrained in an evaporative flow which disperses the disk (see also Gorti \& Hollenbach 2009).\footnote{While Gorti \& Hollenbach (2009) find a qualitatively similar result of decreasing disk lifetime with decreasing dust
opacity (corresponding to lower metallicity in our model), the trend they find is somewhat weaker due to the decreasing efficiency of
photoevaporation driven by the grain photoelectric effect with decreasing dust opacity.}  
Their results closely match the following formula for the disk lifetime $t_{\rm life}$: 

\begin{equation}
t_{\rm life} = 2 \, {\rm Myr} \left(\frac{Z}{{\rm Z}_{\odot}} \right)^{0.77(4-2p)(5-2p)}     \mbox{\ ,}    
\end{equation}
where $p$ is the power-law exponent of the disk surface density profile.  For our fiducial 
case we take $p$ = 0.9, as well as a power-law exponent $q$ of the disk temperature profile of $q$ = 0.6;  
both of these are typical values derived from observations of circumstellar disks (Andrews et al. 2010). 
Thus, we have for our fiducial temperature profile, as a function of the distance $r$ from the central star,

\begin{equation}
    T(r) = 200 \, {\rm K} \left(\frac{r}{1 \, {\rm AU}} \right)^{-0.6}        \mbox{\ ,}
\end{equation}
and for our fiducial surface density profile

\begin{equation}
   \Sigma(r) = 10^3 \, {\rm g \, cm^{-2}} \left(\frac{r}{1 \, {\rm AU}} \right)^{-0.9}         \mbox{\ .}
\end{equation}
These disk parameters are broadly consistent with observed disks (Andrews \& Williams 2007; Andrews et al. 2010), as well as with disk models commonly 
adopted in the literature (e.g. Dullemond \& Dominik 2005; Ercolano \& Clarke 2010).  As discussed in the appendix, we find a weak dependence on these 
particular choices of disk parameters; therefore, for simplicity we shall focus our discussion on this fiducial case.

\subsection{Dust settling timescale}
Here we describe the two processes by which dust grains initially grow in the run-up to planetesimal formation:
dust grain collisions due to settling and collisions due to Brownian motion.  As the process 
of planetesimal formation following the settling of dust grains into the midplane of the disk
is expected to be rapid (e.g. Dullemond \& Dominik 2005; Armitage 2010), we consider the 
longer dust settling timescale to be that which dictates the timescale for planet formation in a disk.  
In Section 6 we briefly describe the impact that other processes affecting the settling timescale have on our results.

\subsubsection{Growth via settling and vice versa}
To model the growth and settling of dust grains we adopt the single particle model of Dullemond \& Dominik (2005).  
In this model dust grains of mass $m$ grow by collisions as they fall through the disk at a velocity $dz/dt$, 
where $z$ is the distance above the plane of the disk, at a rate given by\footnote{We assume that grains only grow via collisions with d$z$/d$t$ $<$ 1 m s$^{-1}$, as collisions at higher velocities do not typcially result in grain sticking (e.g. Blum \& Wurm 2008).}

\begin{equation}
\frac{dm}{dt} = -f_{\rm dg} \rho(z,r) \sigma_{\rm d} \frac{dz}{dt}  \mbox{\ ,}
\end{equation}
where $f_{\rm dg}$ is the dust-to-gas ratio in the disk, $\rho(z,r)$ is the mass density of the disk as a function of $r$ and $z$, and $\sigma_{\rm d}$ is the cross section of the dust grain.
In turn, the velocity at which the dust grain falls through the disk depends on its mass, and is given by (e.g. Dullemond \& Dominik 2005)

\begin{equation}
\frac{dz}{dt} = - \frac{3 \Omega^2_{\rm K} z m}{4 \rho c_{\rm s} \sigma_{\rm d}} \mbox{\ ,}
\end{equation}
where $c_{\rm s}$ is the sound speed of the gas and $\Omega_{\rm K}$ = (G$M_{\rm *}$/$r^3$)$^{\frac{1}{2}}$ is the Keplerian velocity of the disk at a distance $r$ 
from the central star of mass $M_{\rm *}$, which we take to be $M_{\rm *}$ = 0.5 M$_{\odot}$.  This stellar mass is comparable to that assumed by Ercolano \& Clarke (2010) 
in their modeling of disk photoevaporation.\footnote{ Note also that Gorti \& Hollenbach (2009)  find a weak dependence of disk lifetime on stellar mass, for $M_{\rm *}$ $\la$ 3 M$_{\odot}$.}  This is also within the mass range ($\la$ 0.8 M$_{\odot}$) of stars that live for at least a Hubble time and which 
may therefore still be in the Galaxy today even if formed in the early Universe.  Importantly, this allows us to compare our results with data on observed 
metal-poor (and so likely very old) stars, as we do in Section 4. 

The disk is assumed to be in hydrostatic equilibrium such that its density $\rho(z,r)$ is described in terms of its surface density $\Sigma(r)$ by

\begin{equation}
\rho(z,r) = \frac{\Sigma(r)}{h (2 \pi)^{\frac{1}{2}}} {\rm exp}\left(\frac{-z^2}{2h^2} \right) \mbox{\ ,}
\end{equation}
where the scale height of the disk is $h$ = $c_{\rm s}$/$\Omega_{\rm K}$.
Finally, the sound speed at a distance $r$ from the central star is given by $c_{\rm s}(r)$ = (k$_{\rm B}$$T(r)$/$\mu$$m_{\rm H}$)$^{\frac{1}{2}}$,
where $m_{\rm H}$ is the mass of the hydrogen atom, $\mu$ = 2.3 is the mean molecular weight we assume, and $T(r)$ is the temperature 
given by equation (2).

In order to estimate the time it takes for dust grains to settle in the disk, we integrate equations (4) and (5) starting from an initial height of $z$ = 4$h$,
following Dullemond \& Dominik (2005).\footnote{The settling times are not strongly sensitive to the choice of initial height, as the grains 
spend most of their time falling the final scale height or so, as shown in Figures 1 and 2.}  As the initial grain size distribution in metal-poor protostellar disks
in the early Universe is not known (but see e.g. Bianchi \& Schneider 2007; Nozawa et al. 2007; Michalowski et al. 2010; Yamasawa et al. 2011), we consider three different initial grain sizes 
$a_{\rm init}$ in order to bracket the possibilities:  $a_{\rm init}$ = 0.01, 0.1, and 1 $\mu$m.\footnote{Also following Dullemond \& 
Dominik (2005), we assume in our calculations a constant specific weight of dust of 3.6 g cm$^{-3}$ (see also Papaloizou \& Terquem 2006); as these authors note, the settling timescale is not strongly dependent on this choice,
as less dense (more porous) grains will have a larger cross section which makes up for their lower settling velocity.  Related to this, we make the simplifying assumption that the grains are spherical, such that $\sigma_{\rm d}$ = $\pi$$a$$^2$.} 
Finally, to obtain the dependence of the dust settling timescale on the disk metallicity, we vary the value of the dust-to-gas ratio $f_{\rm dg}$,
assuming that it scales linearly with the overall metallicity $Z$ and that solar metallicity corresponds to $f_{\rm dg}$ = 6.5 $\times$ 10$^{-4}$ (e.g. Ercolano \& Clarke 2010). 

In Figure 1, we show the result of our calculation for the three choices of initial dust grain size and for three representative choices of metallicity: $Z$ = 0.025, 0.05, and 0.1 Z$_{\odot}$.
Also shown is the disk lifetime at these three metallicities, given by equation (1).  For a given initial grain size, we can see from this Figure that there exists a minimum metallicity below which 
the disk lifetime is shorter than the grain settling timescale.  It is only for metallicities above this critical value that dust grains will settle into the disk midplane before the 
disk is photoevaporated, thereby allowing planet formation to take place.  For the illustrative cases shown here, with $\Sigma$ = 10$^3$ g cm$^{-2}$ and $T$ = 200 K, even the largest dust grains
(1 $\mu$m initially) will not completely settle before photoevaporation for metallicities $\la$ 0.025 Z$_{\odot}$; however, settling can occur for all initial grain sizes at somewhat higher metallicities
$Z$ $\ga$ 0.05 Z$_{\odot}$.  This already gives us a rough estimate of the critical metallicity for planet formation, but we must consider further processes in order to refine this estimate.

\subsubsection{Growth via Brownian motion}
Irrespective of the velocity at which they settle out of the disk, small scale Brownian motion will cause dust grains to collide and stick together.  
We estimate the rate at which grains grow due to this process as follows.  The relative velocity between of grains with masses $m_{\rm 1}$ and $m_{\rm 2}$ 
is given by

\begin{equation}
v_{\rm rel} = \left(\frac{8 k_{\rm B} T_{\rm dust} (m_{\rm 1} + m_{\rm 2})}{\pi m_{\rm 1} m_{\rm 2}} \right)^{\frac{1}{2}}   \mbox{\ ,}
\end{equation}
where $k_{\rm B}$ is Boltzmann's constant and $T_{\rm dust}$ is the dust temperature, which we have taken to be $T_{\rm dust}$ = 100 K in our fiducial model, 
which is appropriate for relatively low-metallicity disks (see e.g. Clark et al. 2008).  As noted by Dullemond \& Dominik (2005), due to the higher
relative velocity of smaller particles, the coagulation of grains due to Brownian motion quickly establishes a narrow grain size distribution.
Thus, in our single particle model, we can focus on one grain mass $m$ at a time, for which the relative velocity is given simply by

\begin{equation}
v_{\rm rel} \simeq \left(\frac{8 k_{\rm B} T_{\rm dust}}{\pi m} \right)^{\frac{1}{2}}   \mbox{\ .}
\end{equation}
With this, the rate of grain growth is determined by the rate at which grains collide and stick together.  Thus, the 
mass of grains increases according to

\begin{equation}
\frac{dm}{dt} \simeq f_{\rm dg} \rho \sigma_{\rm d} v_{\rm rel}  \mbox{\ ,}
\end{equation}
where $\rho$ is the mass density of the disk and $\sigma_{\rm d}$ is the cross-section of the grain.  As we assume spherical grains for simplicity, this yields 
$\sigma_{\rm d}$ = $\pi$$a^2$, where $a$ is the size of the grain.  Also, as for grain collisions due to settling, we assume grain collisions to lead to growth 
only for velocities $v_{\rm rel}$ $<$ 1 m s$^{-1}$, as collisions at higher velocities typically do not result in sticking (e.g Blum \& Wurm 2008).

In Figure 2 we show the results of our integration of the dust settling equations described in Section 2.2.1 where we have now included
the additional process of grain growth due to Brownian motion.  Comparing Figures 1 and 2, we see that the effect of Brownian motion
is most pronounced for the smallest grains, as expected due to their large relative velocities.  Overall, for intial grain sizes $a_{\rm init}$ $\la$ 0.1 $\mu$m, Brownian motion causes grains to settle to the disk midplane a factor of $<$ 2 times faster than in the case without it.  For our estimation of the critical metallicity for planet formation in the next Section, we account for grain growth via both settling and Brownian motion.

\begin{figure*}[t]
  \centering
  \includegraphics[width=5.in]{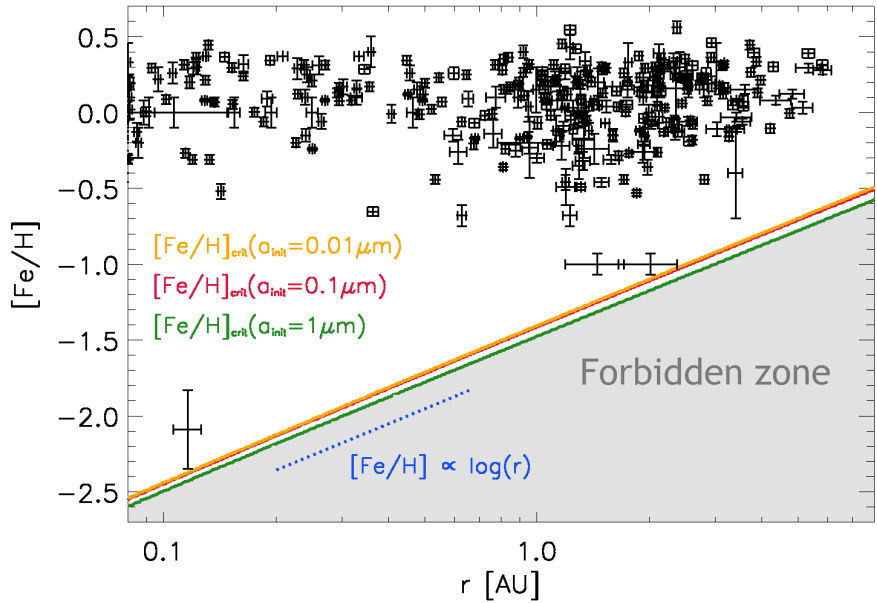}
  \caption
  {The critical metallicity for planet formation, expressed as the iron abundance relative to that of the Sun [Fe/H], 
as a function of distance $r$ from the host star.  The three colored lines correspond to the same initial grain
sizes as shown in Figures 1 and 2, as labeled.  The critical metallicity that we find is not strongly dependent
on the initial grain size distribution, and there is a very close to linear relation between [Fe/H]$_{\rm crit}$ and log($r$) 
as shown by the blue dotted line.  The curves shown here are well approximated by [Fe/H]$_{\rm crit}$ $\simeq$ -1.5 + log($r$/1 AU).
The parameters that we have assumed for our fiducial disk model, as discussed in Section 2.1, are $T(1 $AU$)$ = 200 K, $\Sigma(1 $AU$)$ = 10$^3$ g cm$^{-2}$, $q$ = 0.6, and $p$ = 0.9.
The black crosses show the iron abundance [Fe/H] of observed planet-hosting stars plotted against the semimajor 
axes of the planets' orbits.  
If planetary systems are discovered which lie below this line, in the Forbidden zone shown in gray, it could pose a strong challenge to the core accretion model.
The data shown here are compiled in Wright et al. (2011), except for the lowest metallicity planet at [Fe/H] $\simeq$ -2 which is from Setiawan et al. (2010).}
\end{figure*}

\section{The critical metallicity for planet formation}
To obtain an estimate of the critical metallicity required for planet formation, we compare the two 
timescales presented in the last Section: the lifetime of the disk and the timescale for dust 
grains to settle in the midplane of the disk.  As the latter is required for the initial stage of planetesimal formation to
take place in the core accretion model, the lifetime of the disk must be longer than the settling
timescale in order for planets to form.  The dust-to-gas ratio for
which these timescales are equal provides our estimate of the critical metallicity.

As the dust settling timescale is dependent on the density and temperature of the disk
and because these quantities are functions of the distance $r$ from the host star, the critical
metallicity that we find must itself be a function of $r$ as well.\footnote{In principle the photoevaporation
timescale is also a function of $r$, but as noted by Ercolano \& Clarke (2010) the disk dispersal time is 
much shorter than the overall disk lifetime.  Therefore, we follow these authors and adopt a single disk lifetime
which is independent of $r$.}   For our fiducial disk model given by equations (2) and (3), the curves in Figure 3 show the
critical metallicity at which the disk lifetime is equal to the grain settling timescale as a function of $r$,
for our three representative choices of initial grain size.  Here we have adopted the iron abundance relative to 
solar [Fe/H] as a proxy for the overall metallicity $Z$ and we have once again assumed a linear scaling
between the metallicity and the dust-to-gas ratio.\footnote{Our assumption
that the overall metallicity scales with [Fe/H] should be sound, as at the metallicities we consider ([Fe/H] $\ga$ -2.5) 
the relative chemical abundances inferred in stars tend to be close to the solar values (see e.g. Frebel et al. 2007).}  

As Fig. 3 shows, we find that the cricital metallicity is a strong function of the distance from the host star,
increasing from [Fe/H]$_{\rm crit}$ $\simeq$ -2.5 at $r$ $\simeq$ 0.1 AU to [Fe/H]$_{\rm crit}$ $\simeq$ -0.5 at $r$ $\simeq$ 10 AU.
In fact, the relation that we find for the critical metallicity as a function of $r$ is very close to linear and is
well approximated by 

\begin{equation}
{\rm [Fe/H]}_{\rm crit} \simeq -1.5 + {\rm log}(r / {\rm 1 \, AU}) \mbox{\ .}
\end{equation}
This linear dependence is shown by the blue dotted line in Fig. 3.

We find a very weak dependence of this result on the initial grain size, with [Fe/H]$_{\rm crit}$ decreasing by 
only $\simeq$ 0.1 dex for a two order of magnitude increase in initial grain size.  This confirms that the larger
factor in determining whether the conditions exist for planet formation is indeed the metallicity (more accurately,
the dust-to-gas ratio) of the disk, not the details of the initial grain size distribution.  Furthermore, the result 
that we present here for [Fe/H]$_{\rm crit}$ is not strongly dependent on our assumed model of the disk,
as we show in the appendix.  We therefore conclude that equation (10) provides a simple and relatively
robust estimate of the minimum metallicity for the formation of planets at a distance $r$ in a protostellar disk.
In the next Section, we discuss how this prediction stands up against the available data on observed
planetary systems.

\section{Comparison with data}
Our estimate of the critical metallicity implies a strong prediction, namely that stars hosting planets 
with semimajor axes $r$ should not exhibit metallicities below that given by equation (10) and shown
by the colored curves in Fig. 3.  Shown in gray, we term the region below the lines of critical metallicity
the 'Forbidden zone,' as planet formation should not occur there in the core accretion model.\footnote{This terminology is adopted from Frebel et al. (2007) who use it 
to describe the low-metallicity regime in which low-mass stars should not be found as predicted by the atomic cooling theory of their formation (Bromm \& Loeb 2003). } 
We now turn to test this prediction using observational data on planetary systems from the literature.  

In Fig. 3 we show the [Fe/H] values inferred for $\simeq$ 320 planet-hosting stars reported in
the literature and compiled by Wright et al. (2011)\footnote{We have taken these data directly from exoplanets.org (Wright et al. 2011).}, plotted against the semimajor axes of the 
orbits inferred for the planets they host.  In addition we have also plotted here the [Fe/H] and semimajor axis values inferred 
for the very metal-poor ([Fe/H] $\simeq$ -2) planetary system reported by Setiawan et al. (2010).  As shown
in the Figure, the vast majority of the data are for planetary systems with metallicities greatly exceeding 
the minimum value that we predict.  However, the three lowest metallicity systems, with [Fe/H] $\simeq$ -1 (Niedzielski et al. 2009)
and [Fe/H] $\simeq$ -2 (Setiawan et al. 2010), exhibit metallicities only slightly greater than the critical value.
Therefore, while the data shown here are not in conflict with our predicted critical metallicity, it would be of great interest
to test the theory with more data on metal-poor planetary systems.  As we will discuss further in Section 6, 
planetary systems which are found to lay below the critical lines shown in Fig. 3 could
strongly challenge the core accretion model of planet formation.  If they are found, we may have to 
revisit this popular model for the formation of, at least, low-metallicity planets.  We note that this may 
be best accomplished by microlensing surveys, given the sensitivity of this technique to planets 
at large distance from their host stars (see e.g. Cassan et al. 2012), where the critical metallicity is highest.

It is interesting to also test the critical metallicity given by equation (10) against the data on our own Solar System.
While not shown in Fig. 3, the most distant planet in the Solar System, Neptune, lies just above the Forbidden zone as well, at $r$ $\simeq$ 30 AU and [Fe/H] = 0.  The dwarf planet Pluto,
however, lies slightly within the Forbidden zone; as with the Kuiper belt objects, Pluto's wide orbit may be due to gravitational interaction with the outer gas giants 
and may not reflect the location at which it formed (see e.g. Levison et al. 2008).
Furthermore, it is intriguing to note that, as discussed by Levinson et al. (2008), the popular {\it Nice} model which 
explains the orbits of the gas giants (e.g. Tsiganis et al. 2005) requires the population of planetesimals 
in the Solar nebula to have extended only out to $\sim$ 30 - 35 AU, which is exactly what we obtain from equation (10) for the maximum
distance at which planetesimal formation is possible at solar metallicity.  At larger distances from the Sun, the dust settling timescale
may have been longer than the lifetime of the disk from which the Solar System formed, thus providing a natural explanation for the
extent of the protoplanetary disk that is inferred from the {\it Nice} model.

We note that we have made a number of assumptions in order to make this comparison with the data.  
The first is that the surface metallicity of the host star is the same as that of the protostellar disk from which it and its planets formed.
This is likely to be valid, although in principle stars may accrete some metals from the ISM during 
their lives (but in most cases not enough to alter their surface composition; see e.g. Frebel et al. 2009; Johnson \& Khochfar 2011) 
and may otherwise alter their surface abundances due to mass loss or convection. 

The second assumption is that the planets shown here have roughly circular orbits.  If the orbits are instead highly eccentric,
then it becomes difficult to compare the data to our theoretical prediction which is calculated at a single distance $r$ from the host star.
Fortunately, the orbital eccentricity of the three lowest metallicity planets shown in Fig. 3 is modest ($\simeq$ 0.2; see Setiawan et al. 2010;
Niedzielski et al. 2009).  Indeed, the vast majority of the planets shown here have orbital eccentricities $\la$ 0.5 (see Wright et al. 2011 and references therein). 
Therefore, on this account these data should indeed allow for a fairly reliable comparison of data and theory.  

The third assumption that we make is that the planets shown here have not migrated from their 
place of birth in the disk.  While we do not venture to address the validity of this assumption, we do 
note that planets are believed to commonly migrate inwards toward their host stars (e.g. Papaloizou \& Terquem 2006), although outward movement
can occur via interaction with the disk (see e.g. Hahn \& Malhotra 1999) or through gravitational scattering (e.g. Veras et al. 2009; see also Perets \& Kouwenhoven 2012).   
If inward migration occurs, this implies that the observed semimajor axes shown in Fig. 3 are lower limits
for the semimajor axes of the planets at the time of their formation.  This could place
some planets in the Forbidden zone, in conflict with our predicted lower metallicity limit. 
Conversely, if outward migration occurs, this could potentially place observed planets in the Forbidden zone even if they reside safely within
it at the time of their formation.\footnote{There have been detected several exoplanets with very wide orbits ($r$ $\ga$ 100 AU; e.g. Lafreni{\' e}re et al. 2008, 2011; Biller et al. 2011; Ireland et al. 2011) which may have originally formed closer to their host stars (e.g. Veras et al. 2009; Perets \& Kouwenhoven 2012) or may have formed from gravitational instabilities in the disk (e.g. Dodson-Robinson et al. 2009).  We note, however, that planets 
at these large radii could still lie above the Forbidden zone if their host stars have [Fe/H] $\ga$ 0.5.  It is also very likely that the
planet orbiting a low-metallicity ([Fe/H] $\sim$ -1.3) binary star system at $r$ $\sim$ 23 AU reported by Sigurdsson et al. (2003) was originally 
formed much closer to its host star.}   
We further discuss these processes, along with others that may complicate our results, in Section 6.

\section{Implications for the first planets and life}
Here we discuss the implications of the results presented in Sections 3 and 4.  We first 
describe the likely properties of the first planets as implied by our estimate of
the critical metallicity for planet formation.  Then, in the following Section, we turn to discuss how our findings
may pose a challenge to the core accretion model.

\subsection{Properties of the first planets}
We have found that the early, planetesimal formation stage in the core accretion scenario of planet formation can only take place once 
a minimum metallicity has been generated in protostellar disks. 
While Population (Pop) III stars thus could not have hosted planets formed via core accretion, it is expected
that the supernovae of these first stars ejected the metals and dust from which planets may have 
formed around Pop II stars (e.g. Heger \& Woosley 2002; Schneider et al. 2004; Nozawa et al. 2007; Cherchneff \& Dwek 2010).
Cosmological simulations of Pop III supernova explosions suggest that the gas is enriched 
to metallicities of up to $\simeq$ 10$^{-3}$ Z$_{\odot}$ when it recollapses to form second generation stars (e.g. Wise \& Abel 
2008; Greif et al. 2010; but see Whalen et al. 2008 on the case of relatively low-mass Pop III supernovae).  
This metallicity is above that required for the formation of the first low-mass Pop II stars which may still
be alive today (e.g. Bromm et al. 2009), and which may in principle host planets.  
Here we briefly discuss what our findings suggest would be the properties of these planets.

Following equation (10), we can estimate the distance within which planets may have formed around 
these first low-mass Pop II stars.  Assuming a metallicity $Z$ $\simeq$ 10$^{-3}$ Z$_{\odot}$,
as mentioned above,  we find this to be $r$ $\la$ 0.03 AU
Therefore, if planets did form around the earliest metal-enriched stars the core accretion model predicts that 
they must have formed with very compact orbits.  

We can obtain a rough upper limit for the mass of a planet formed around a star of a given metallicity $Z$,
by taking the mass budget for the planet to be all the material in the disk interior to the largest possible semimajor axis $r_{\rm max}$
at which planet formation could occur.  This is obtained by
inverting equation (10) as

\begin{equation}
r_{\rm max} \simeq 32 \, {\rm AU} \, \left(\frac{Z}{{\rm Z_{\odot}}} \right)\mbox{\ ,}
\end{equation}
where we have again assumed the iron abundance [Fe/H] (and the dust-to-gas ratio) to scale with metallicity $Z$.
Integrating over the surface density profile out to this radius then yields the upper mass limit $M_{\rm max}$ of the planet as

\begin{eqnarray}
M_{\rm max} & \simeq & \int^{r_{\rm max}}_{0} 2 \pi r \Sigma(r)  dr \\ \nonumber
& \simeq & 10^4 \, {\rm M_{\earth}} \, \left(\frac{\Sigma(1 {\rm AU})}{10^3 \, {\rm g \, cm^{-2}}} \right) \left(\frac{Z}{{\rm Z_{\odot}}}\right)^{1.1} \mbox{\ ,}
\end{eqnarray}
where we have used the general form for the surface density profile described in Section 2.1.  Here $\Sigma(1$AU$)$ is the surface density of the disk at 1 AU and
we have assumed a surface density profile power-law exonent $p$ = 0.9 as in our fiducial disk model. 
Equation (12) shows that the maximum mass of a planet is a strong function of the metallicity of the gas from which it forms.  In particular, it shows
that the mass available for the first planets may be only $\sim$ 10 M$_{\earth}$, again assuming our fiducial disk model and a metallicity $Z$ $\simeq$ 10$^{-3}$ Z$_{\odot}$ for the first metal-enriched stars.\footnote{We note also that at solar metallicity Z$_{\odot}$ the mass budget for planets given by equation (12) is consistent with the total mass of planets in the Solar System, which is $\simeq$ 470 M$_{\earth}$.}  

However, given the close proximity of these planets to their host stars, $r$ $\la$ 0.03 AU as given above, 
the temperature of the disk may be high enough to sublimate dust grains, thereby strongly suppressing grain growth and planetesimal formation.
Using our fiducial model for the temperature profile of the disk, we can estimate the distance from the host star within which dust submlimation will be important.  
At $r$ $\simeq$ 0.03 AU, we find the temperature of the disk to be $T$ $\simeq$ 1600 K, which
is indeed well above the sublimation temperature of many forms of dust (e.g. Kobayashi et al. 2011).
We note also that the host star may easily heat the material to temperatures high enough to evaporate the lightest elements out of the atmosphere of any planet that does form.  
Especially at low-metallicity, the vast majority of the gas is hydrogen and helium, and heavy elements compose only a fraction $\sim$ 10$^{-5}$ of the total mass, 
for example, at $Z$ $\sim$ 10$^{-3}$ Z$_{\odot}$.  
It thus appears that planet formation around the first low-mass metal-enriched stars may have been strongly suppressed due to proximity to the host star.
Any planets that did form would likely have been very low mass, in addition to having very compact orbits.

\subsection{The first Earth-like planets and life}
While we have found that planets formed around the first low-mass metal-enriched stars were likely to have been too small and hot to host life, we can use equation (12) to estimate the 
metallicity at which the formation of Earth-like planets first becomes possible.  To do this, we note that, as an Earth-like planet must be 
composed almost entirely of heavy elements, there must be at least an Earth mass in metals available in order for one to form.  Multiplying the right side of
equation (12) by the metallicity thus yields the following maximum mass of a planet composed of elements heavier than helium,
as a function of metallicity:

\begin{equation}
M_{\rm max} \simeq 200 \, {\rm M_{\earth}} \, \left(\frac{\Sigma(1 {\rm AU})}{10^3 \, {\rm g \, cm^{-2}}} \right) \left(\frac{Z}{{\rm Z_{\odot}}}\right)^{2.1} \mbox{\ ,}
\end{equation}
where we have assumed the solar metallicity to be Z$_{\odot}$ = 0.02, and we note again that here we have assumed our fiducial value of $p$ = 0.9 for the disk surface density power-law exponent.  Solving this for the metallicity at which the maximum mass of heavy elements available is an 
Earth mass, we find that for an Earth-like planet to form the disk must have a metallicity 

\begin{equation}
Z \ga 0.1 \, {\rm Z}_{\odot} \, \left(\frac{\Sigma(1 {\rm AU})}{  10^3 \, {\rm g \, cm^{-2}}}\right)^{-0.48} \mbox{\ .}
\end{equation}

In our fiducial disk model, due to the steep ($p$ = 0.9) density profile of the disk the bulk of the material available for planet formation will lie near $r_{\rm max}$.  Thus, using this minimum metallicity in
equation (11) shows that  these first Earth-mass planets would have likely formed at radii $r$ $\sim$ 3 AU, assuming our fiducial value of $\Sigma(1 {\rm AU})$ = 10$^3$ g cm$^{-2}$.
In turn, this implies that such a planet would lie within the habitable zone
of its host star only if the star had a mass $\ga$ 1.5 M$_{\odot}$ (e.g. Kasting et al. 1993).\footnote{We note, however, that such a high 
mass star may also exhibit stronger radiative feedback on the disk than we have assumed following Ercolano \& Clarke (2010); 
if so, then the critical metallicity we've assumed may be an underestimate and an Earth mass planet may form at somewhat smaller distance $r$ 
than we have estimated here.}  
Thus, while the first Earth mass planets could have formed only at relatively high metallicities ($Z$ $\ga$ 0.1 Z$_{\odot}$),\footnote{Interestingly, this is exactly in the range of minimum metallicities required for Earth-like planets that has been assumed by previous authors, also based on the mass budget of heavy elements in the disk (e.g. Lineweaver 2001; Lineweaver et al. 2005; for somewhat higher estimates see e.g. Gonzalez et al. 2001; Zinnecker 2004).} 
it may be that life could only take hold on these planets if their host stars were sufficiently massive ($\ga$ 1.5 M$_{\odot}$).
As such high mass stars live $\la$ 4 Gyr, much less than a Hubble time, it may be the case that the first life in the Universe
came and went long ago.  In order for life to evolve on Earth-like planets orbiting more long-lived (lower mass) stars, the 
metallicity of the disk would have to be higher than the minimum value given by equation (14).  Indeed, this is consistent with the formation of the Earth 
itself at solar metallicity.

We note furthermore, however, that the formation of Earth-like planets is not itself a sufficient prerequisite for life to take hold.  
In particular, early galaxies were likely rife with supernovae and may have hosted actively growing central black holes, 
both strong sources of life-threatening radiation (see e.g. Clarke 1981; Crutzen \& Bruhl 1996; Lineweaver et al. 2005; Ejzak 
et al. 2007).  This provides further reason to expect that the conditions for life emerged only after the earliest epoch of galaxy formation.

\section{A challenge to the core accretion model?}
While the critical metallicity that we have estimated in the context of the early, dust coagulation stage of the core accretion model 
of planet formation is in agreement with the available data, we emphasize that the our estimate
is a strong lower limit.  There are a number of assumptions in our 
calculation that, if relaxed, may place the data shown in Figure 3 in conflict with the core accretion 
model.  These are the following: 

(1) We have assumed that the dust-to-gas ratio scales linearly with [Fe/H].  However, this may not
be the case, and it has been suggested that the dust-to-gas ratio may decrease faster than this with 
decreasing metallicity (see e.g. Inoue 2003, 2011).  If so, then this would require an upward revision of 
the critical metallicity that we have found and an expansion of the Forbidden zone shown in Fig. 3.

(2) We have not accounted for the destruction of dust in the disk, which is likely to occur to 
some degree (e.g. Dullemond \& Dominik 2005).  As mentioned in Section 5.1, this is especially true at small distances (e.g. $r$ $\la$ 1 AU) from 
the host star where the temperature is high enough to sublimate many forms of dust (see e.g. Todini \& Ferrara 2001; Kobayashi et al. 2011).
If dust destruction is efficient, it is difficult to explain how grains grow sufficiently rapidly at low metallicity to settle into the midplane 
and form planetesimals.

(3) As mentioned in Section 4, the observed planets shown in Fig. 3 may have migrated 
inwards from their birth sites.  If they did form much further out in the disk, our model may not 
be adequate to explain how they formed in the low density outskirts of the disk before the disk 
was photoevaporated.\footnote{This is likely to be the case especially for the most metal-poor ([Fe/H] $\simeq$ -2) planetary
system.   If this planet, with a mass $\ga$ 1.25 times that of Jupiter, was only able to grow through accretion of the gas within its current orbit 
the disk surface density must have been $\ga$ 3 $\times$ 10$^5$ g cm$^{-2}$; this is very high and suggests that the planet instead 
grew by accretion at larger radii and then migrated inward.}  

(4) We have also not accounted for the inward radial drift 
of dust that may occur during settling (e.g. Armitage 2010; but see Birnstiel 2011) or for the 
possibility that dust grains are entrained in an inward propagating disk accretion flow during settling.
These processes, too, if effective would require an upward revision of the critical metallicity that 
may place the data in the Forbidden zone in Fig. 3.  

(5) We have not included the effect of turbulence in our modeling.  As discussed by, e.g., Armitage et al. (2010)
turbulence can strongly resist dust settling (see also e.g. Birnstiel et al. 2012).  Therefore, even weak turbulence 
could raise the critical metallicity substantially above what we show in Fig. 3.  Previous studies provide evidence 
that the critical metallicity may indeed be higher in turbulent disks in which planetesimals form via streaming 
(Johansen et al. 2009) and secular gravitational (Takeuchi \& Ida 2012) instabilities.  Also, the lower abundance of dust grains 
at lower metallicity may lead to a higher ionization fraction in the disk (e.g. Armitage 2011; Bai 2011), which would likely result in stronger turbulence driven by magnetohydrodynamic instabilies at lower metallicity.

Given that we may clearly be underestimating the critical metallicity, the fact that the three planets with [Fe/H] $\la$ 0.1 
discovered to date exhibit properties placing them just outside the Forbidden zone suggests that
these planets may already pose a serious problem for the core accretion model.  Furthermore, we note that all 
three of these planets have masses exceeding that of Jupiter, and in the case of the two planets at [Fe/H] $\simeq$ -1 they have
masses at least 10 times that of Jupiter.  However, at the lowest metallicities the core accretion model would predict that planets should exhibit
relatively low masses, as they have little time after dust settling and planetesimal formation
to accrete mass from the disk (e.g. Pollack et al. 1996; Ida \& Lin 2005); indeed, this expectation is consistent with the weak dependence of 
the frequency of relatively low-mass ($\la$ 30 M$_{\earth}$) planets on metallicity (for [Fe/H] $\ga$ -0.4) reported by Mayor et al. (2011).  
The reported detections of relatively massive planets at the lowest metallicities are not expected in the core accretion model.  

Nevertheless, the data shown in Figure 3 are not in violation of the critical metallicity that we have found from our simple modeling.  
If future searches for planets around metal-poor stars fail to find planetary systems that are in violation of this critical metallicity,
this may imply that the core accretion is in fact sound.  In this event, even to be consistent with the existing data, the assumptions listed above
would likely have to hold.  This would imply the following: 
(1) dust-to-metals ratios in low-metallcity disks are not significantly lower than we have assumed for the solar system; (2) dust destruction does not
substantially slow the growth and settling of grains; (3) inward migration of planets does not strongly affect the semimajor axes of low-metallicity planets;
(4) Radial drift and large scale accretion flows do not carry dust grains large distances inward through disks during settling; and (5) turbulence does 
not strongly stir up dust grains and prevent settling.

\section{Conclusions}
We have estimated the minimum metallicity required for planet formation in the popular core accretion model,
by comparing the time required for dust to settle into the midplane of a circumstellar disk to the disk lifetime.
Only for sufficiently high dust-to-gas ratios is there time for dust settling, which precedes planetesimal formation, 
to occur before the disk is photoevaporated by the host star.  Assuming that the dust-to-gas ratio scales with the 
metallicity of the gas at low metallicities, we find a critical metallicity for planet formation that is a strong function of the distance from 
the host star.  In our fiducial model, which assumes typical properties of observed disks, we find
[Fe/H]$_{\rm crit}$ $\simeq$ -1.5 + log($r$/1 AU).  This result, however, is not strongly dependent 
on our choice of disk parameters, as we show in the appendix.

We find that this prediction of the critical metallicity is in agreement with the available data, and that 
it provides a natural explanation for the extent of the Solar protoplanetary disk as inferred from 
the popular {\it Nice} model which explains the orbits of the gas giants.
That said, there have been discovered low-metallicity planetary systems which exhibit 
metallicities just above the critical value we find which, as discussed in Section 6, is a 
strong lower limit for the critical metallicity based on the core accretion model.  
The results of future searches for planets around metal-poor stars will therefore be of great interest, as if planets are found
in the Forbidden zone shown in Fig. 3, this will challenge the core accretion model.  Given that the critical metallicity is highest 
at large distances from the host star, microlensing surveys that 
are sensitive to planets with large semimajor axes may be best suited for this task.  Indeed, the few exoplanets that 
have been found with very wide orbits ($r$ $\ga$ 100 AU; e.g. Lafreni{\' e}re et al. 2008, 2011; Biller et al. 2011; Ireland et al. 2011) 
may already pose a challenge to the core accretion model, and it has been suggested that the gravitational instability model (e.g. Boss 1997) 
may better explain their formation (e.g. Dodson-Robinson et al. 2009).  Alternatively, these planets may have originally formed 
via core accretion closer to their host stars (e.g. Veras et al. 2009; Perets \& Kouwenhoven 2012).  We note, however, that planets 
at these large radii could still lie above the Forbidden zone shown in Fig. 3 if their host stars are found to have [Fe/H] $\ga$ 0.5.

If the core accretion model does correctly describe planet formation, then we have found that our estimate 
of the critical metallicity implies that if planets did form around the first low-mass metal-enriched stars 
they would have been extremely low mass and would have had very compact orbits.   The first Earth-like planets, in turn, likely formed around stars
with relatively high metallicities ($\ga$ 0.1 Z$_{\odot}$) and may have resided in the habitable zones of their host stars only if these stars
had sufficiently high masses ($\ga$ 1.5 M$_{\odot}$).  If life did evolve on such early Earth-like planets, it may therefore
have since been extinguished with the death of the host star, which at these super-solar masses would live less than 4 Gyr, a small fraction 
of the age of the Universe.

In the past decade there has been a lively debate with regard to the processes that set the critical metallicity
for the first low-mass stars in the Universe.  As discussed by e.g. Frebel et al. (2007), the two competing theories, those based on 
atomic line cooling and dust cooling respectively, are best tested with observations of metal-poor stars.  Indeed, this approach has
led recently to a strong challenge to the atomic line cooling theory (Caffau et al. 2011, 2012), implying that it is perhaps
instead dust cooling which facilitates the formation of the first low-mass stars (e.g. Klessen et al. 2012).
If the progress of the low-metallicity planet hunting community follows a track analogous to that 
of the low-metallicity star hunting community, we can expect future data to provide strong tests of
the core accretion model, and in turn to reveal more about the the nature of the first planets in the Universe.

\section*{Acknowledgements}
We gratefully acknowledge the support of the U.S. Department of Energy through the LANL/LDRD Program for this work.
JLJ also gratefully acknowledges the support of a LDRD Director's Postdoctoral Fellowship at Los 
Alamos National Laboratory.  We are thankful to Volker Bromm, Pawan Kumar, Doug Lin, Alexia Schulz, and Tsing-Wai 
Wong for helpful discussions and feedback, and to the anonymous reviewer for a thorough and constructive report.  
JLJ also thanks Claudio Dalla Vecchia for help with IDL color tables.  This research 
has made use of the Exoplanet Orbit Database and the Exoplanet Data Explorer at exoplanets.org.

\bibliographystyle{apj}


\appendix

\section{Dependence on disk parameters}
Here we discuss the relatively weak dependence of our estimate of the critical metallicity on the properties of the
disk.  As described in Section 2.1 our disk models are characterized by four quantities: the temperature 
gradient parameter $q$, the density gradient parameter $p$, and the normalizations of the temperature
and density profiles.  These parameters impact our estimates of the settling timescale and the 
disk lifetime in non-trivial ways, following equations (1), (4), (5), (6) and (9).  
The distance dust must fall to reach the disk midplane is set by the scale height, which increases with temperature.
The dust settling velocity decreases with both temperature and surface density, but increases with grain mass.  Also,
the rate at which grains grow in mass increases with grain mass and distance above the disk, but decreases with temperature.
Finally, the disk lifetime is dependent on the surface density profile parameter $p$.

\begin{figure*}[t]
  \centering
  \includegraphics[width=6.5in]{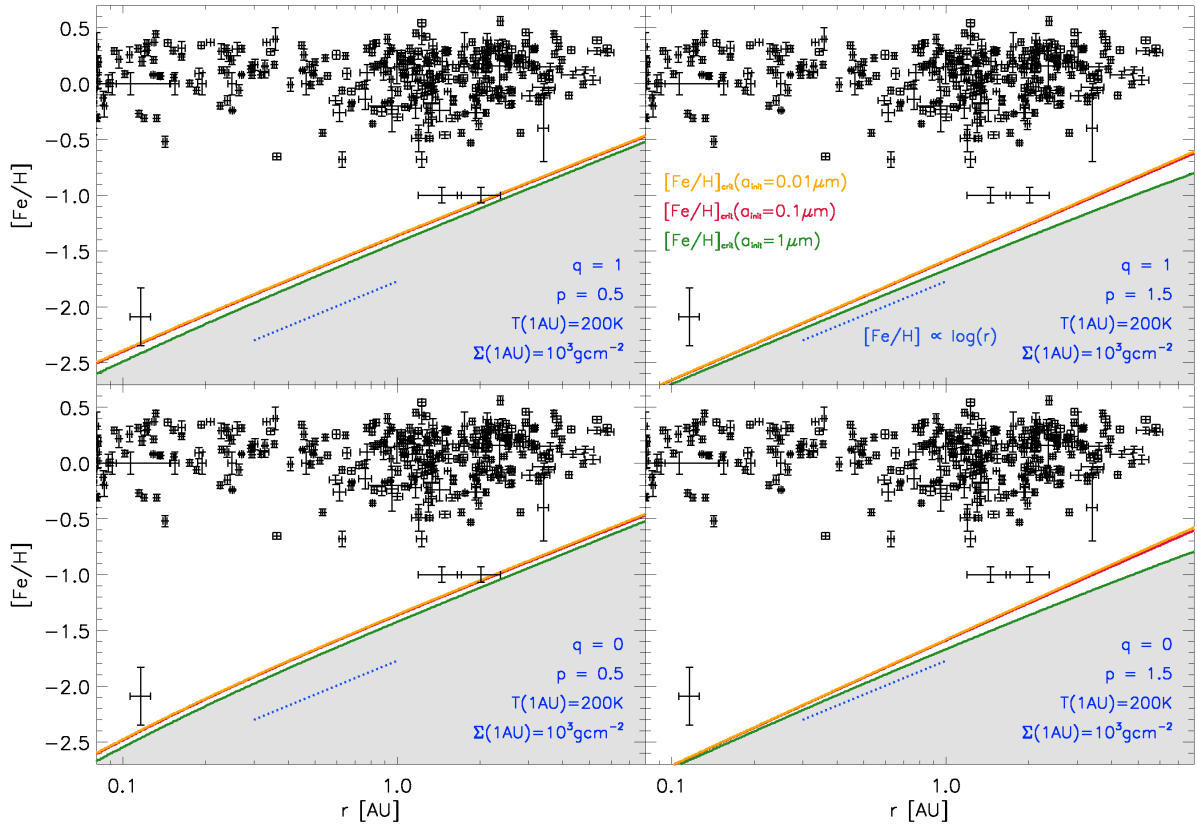}
  \caption
  {The variation of the critical metallicity with different disk surface density and temperature profiles. These four plots are all the same 
as Fig. 3 except that here we have chosen different values of $q$ and $p$, the temperature and surface density power-law exponents, in each panel 
as shown in blue.  Also shown are the temperature and surface density normalizations at $r$ = 1 AU, which here are the same as in Fig. 3.
While our choices of gradient parameters here broadly bracket those of observed disks, we find that the curves of critical metallicity
do not strongly depend on these choices, varying by only a factor of $\simeq$ 2 between the plots.  Also, as shown by the blue dotted lines,
in all cases the critical metallicity has nearly the same dependence on distance from the host star, i.e. [Fe/H]$_{\rm crit}$ $\propto$ log($r$).
As in Fig. 3, the gray region highlights the 'Forbidden zone' where planets should not be found, according to the core accretion model.}
\end{figure*}

\begin{figure*}[t]
  \centering
  \includegraphics[width=6.5in]{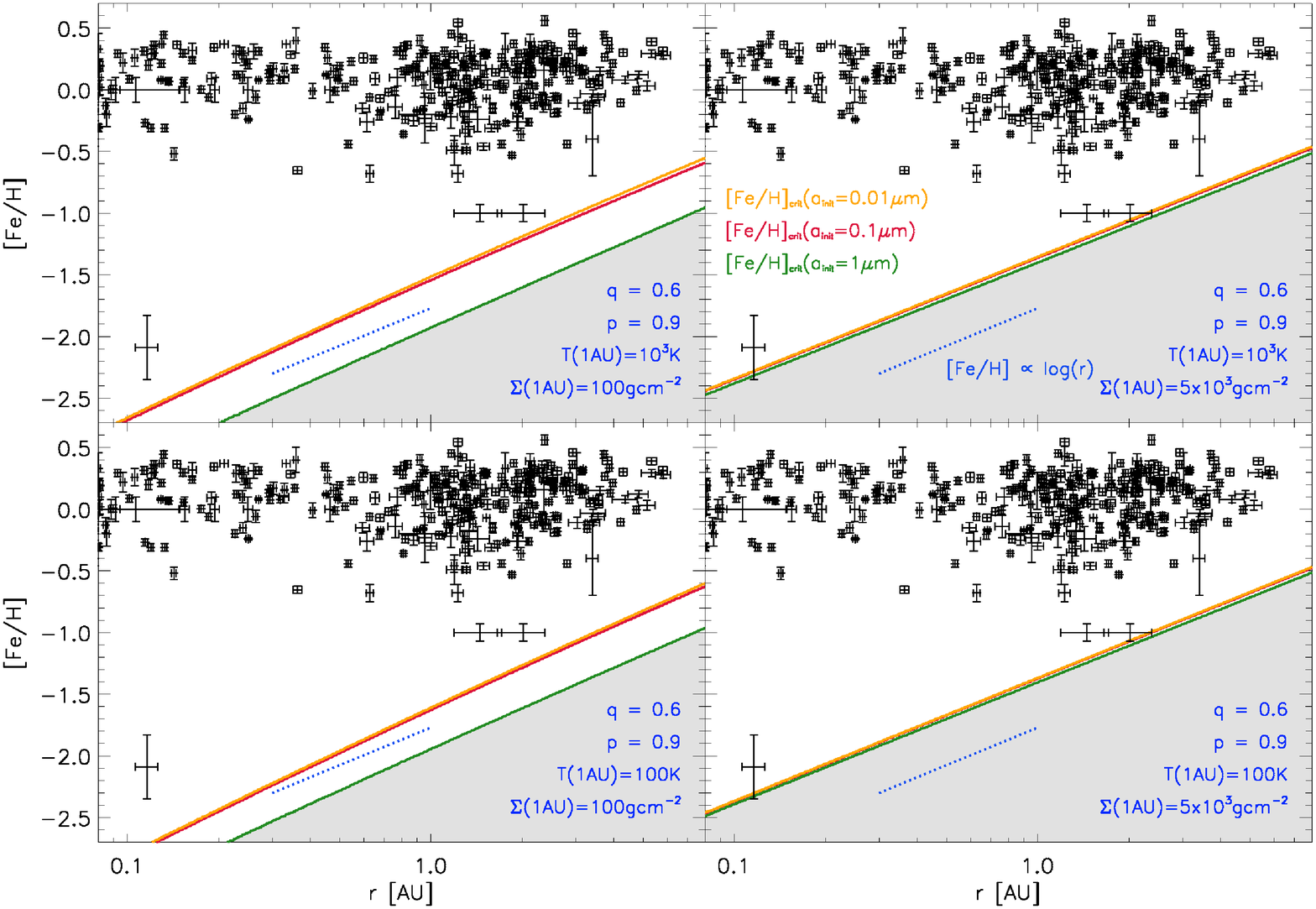}
  \caption
  {The same as Figure A4, but now the temperature and surface density normalizations at $r$ = 1 AU are varied between the panels while the gradient parameters $q$ and $p$
are the same, as shown in blue. While the temperature normalization has very little effect on the curves of critical metallicity, between the two surface density normalizations the
critical metallicity varies by a factor of $\simeq$ 3.  Still, the dependence on $r$ is still very close to [Fe/H]$_{\rm crit}$ $\propto$ log($r$), and the curves are also consistent
with the data in all cases.}
\end{figure*}

To check how sensitive our results are on the disk properties, we plot again the critical metallicity
as a function of $r$ along with the data as in Fig. 3.  In Figure A4, we show four plots with the same
temperature and surface density normalizations at $r$ = 1 AU (as we have chosen for our fiducial model shown in Fig. 3), 
but each with different combinations of the $p$ and $q$ gradient parameters.  We vary the temperature gradient parameter $q$ between
0 and 1, which broadly brackets the values inferred for observed circumstellar disks (see Andrews \& Williams 2007; Laibe et al. 2011).
We vary the surface density gradient parameter $p$ between 0.5 and 1.5, which also brackets the values
inferred for circumstellar disks (see Andrews et al. 2010).   As the plots show, the resulting critical metallicity 
curves all still exhibit roughly the same dependence on $r$, namely they are still well described by [Fe/H]$_{\rm crit}$ $\propto$ log($r$) 
as in our fiducial model.  Also the normalization of the curves varies relatively little between the plots, 
being in all of them within a factor of $\simeq$ 2 of that given by equation (10).  

In Figure A5, we maintain the same gradient parameters as in our fiducial model, but now we vary
the normalizations of the surface density and temperature.  We vary the temperature at $r$ = 1 AU between
100 K and 10$^3$ K, while we vary the surface density normalization at $r$ = 1 AU between 100 g cm$^{-2}$
and 5 $\times$ 10$^3$ g cm$^{-2}$.  These choices broadly bracket the values reported by Andrews et al. (2010; see also e.g. Kuchner 2004; Davis 2005) 
for observed circumstellar disks.  While the plots show that the dependence on the 
disk temperature normalization is very small, the dependence on the surface density normalization is larger,
the critical metallicity varying by a factor of $\simeq$ 3 between the plots.  Nevertheless, the curves of 
critical metallicity still roughly follow the [Fe/H]$_{\rm crit}$ $\propto$ log($r$) trend as found in our fiducial model.  Also, 
in all cases shown in both Figs. A4 and A5 the data are consistent with our 
simple calculation of the critical metallicity.

Furthermore, we note that the disk lifetime $t_{\rm life}$ given by equation (1) that we have used here should, 
in principle, be dependent on the normalization of the disk density profile, in addition to the surface density 
power-law exponent $p$ and the metallicity $Z$.  As a higher surface density normalization will lead to higher 
column density, the disk will be more shielded from X-ray radiation and the lifetime $t_{\rm life}$ should thus 
increase (see Ercolano \& Clarke 2010); in turn, this would lead to a lower critical metallicity.  Therefore, for the higher (lower) density 
normalizations we've chosen here, the critical metallcity may be somewhat lower (higher) than shown.
Accounting for this effect would thus bring the curves shown here in better agreement with the fiducial curves
shown in Fig. 3, suggesting that the critical metallicity we calculate is likely to vary by even less than the 
factor of $\simeq$ 3 shown in Fig. A5.

We conclude that, not only is our result for the critical metallicity not strongly dependent on the initial 
dust grain size distribution, it is also not strongly dependent on the properties of the disk.

\end{document}